\documentclass[aps,prl,reprint,groupedaddress,nofootinbib]{revtex4-1}
\usepackage{hyperref}
\usepackage{amsmath}
 \usepackage{multirow}
\usepackage{array}
\usepackage{url}
\newcolumntype{L}[1]{>{\raggedright\let\newline\\\arraybacksslash\hspace{0pt}}m{#1}}
\newcolumntype{C}[1]{>{\centering\let\newline\\\arraybackslash\hspace{0pt}}m{#1}}
\newcolumntype{R}[1]{>{\raggedleft\let\newline\\\arraybackslash\hspace{0pt}}m{#1}}
\usepackage{float}
\usepackage{graphicx}
\usepackage{epsfig}
\usepackage{psfrag}
\usepackage{color}
\usepackage{slashed}

\usepackage{amsfonts}
\usepackage{amssymb}
\usepackage{tikz}
\usepackage{tikz}
\usetikzlibrary{positioning,arrows}
\usetikzlibrary{decorations.pathmorphing}
\usetikzlibrary{decorations.markings}

\newcommand*{\be}{\begin{equation}}
\newcommand*{\ee}{\end{equation}}
\newcommand*{\bea}{\begin{eqnarray}}
\newcommand*{\eea}{\end{eqnarray}}

\newcommand{\comment}[1]{}

\newcommand{\cref}[1]{Chapter~\ref{c.#1}}

\def\beq{\begin{equation}}
\def\eeq{\end{equation}}
\def\bea{\begin{eqnarray}}
\def\eea{\end{eqnarray}}
\def\ba{\begin{array}}
\def\ea{\end{array}}
\def\bi{\begin{itemize}}
\def\ei{\end{itemize}}
\def\be{\begin{enumerate}}
\def\ee{\end{enumerate}}
\def\bc{\begin{center}}
\def\ec{\end{center}}
\def\bt{\begin{table}}
\def\et{\end{table}}
\def\btb{\begin{tabular}}
\def\etb{\end{tabular}}

\def\lsim{\raise0.3ex\hbox{$\;<$\kern-0.75em\raise-1.1ex\hbox{$\sim\;$}}}
\def\gsim{\raise0.3ex\hbox{$\;>$\kern-0.75em\raise-1.1ex\hbox{$\sim\;$}}}
	
\begin{document}

\title{ Confronting  $B$ anomalies with low energy  parity violation}

\author{G. D'Ambrosio}
\affiliation{INFN-Sezione di Napoli, Via Cintia, 80126 Napoli, Italia}
\author{ A. M. Iyer}
\affiliation{INFN-Sezione di Napoli, Via Cintia, 80126 Napoli, Italia}
\author{   F. Piccinini}
\affiliation{	INFN, Sezione di Pavia, via A. Bassi 6, 27100 Pavia – Italy}
 \author{A.D. Polosa}
\affiliation{Dipartimento di Fisica and INFN, Sapienza Universit$\grave{a}$ di Roma, P.le Aldo Moro 2, I-00185 Roma, Italy}
\begin{abstract}
	Indirect searches have the potential to probe scales beyond the realm of direct searches. In this letter we consider the implications  of two parity violating experiments: weak charge of proton $Q_W^p$ and the Caesium atom $Q_W^{Cs}$ on the solutions to lepton flavour non-universality violations (LFUV) in the decay of $B$ mesons. 
	Working in a generic implementation of a minimal $Z^\prime$ model, we assume the primary contribution being due to the electron to facilitate comparison with the low $q^2$ parity violating experiments. We demonstrate that the conclusion is characterized by different limiting behavior depending on the chirality of the lepton current. The correlation developed in this study demonstrates the effectiveness in studying the synergy between different experiments leading to a deeper understanding of the interpretation of the existing data. It is shown that a possible future improvement in the parity violating experiments can have  far reaching implications in the context of direct searches. We also comment on the prospect of addition of the muon to the fits and the role it plays in ameliorating the constraints on models of $Z'$. This offers a complimentary understanding of the pattern of the coupling of the NP to the leptons, strongly suggesting either a muon only or a combination of solutions to the anomalies.

\end{abstract}

\maketitle

Indirect measurements are extremely sensitive to small deviations from the Standard Model (SM) predictions. These could be in the form of flavour violating transitions or an estimate of flavour diagonal effects. New physics, if present, must have different but correlative implications for physics measurements across different energy scales. This serves as a useful motivation to explore the combined effects of different probes on simplified extensions to the SM. 

Semi leptonic decays of the $B$ mesons constitutes one of the strongest hints for beyond standard model (BSM) physics. The measurement of the theoretically clean  ratio $R_K=\mathcal{B}(B^+\rightarrow K^+\mu^+\mu^-)/\mathcal{B}(B^+\rightarrow K^+e^+e^-)$ \cite{Aaij:2019wad} 
%\begin{eqnarray*}
$	R_K\vert_{q^2=1-6~GeV^2}=0.846^{+0.060}_{-0.0 16}~(stat)\pm 0.014~(syst)$
	\label{ratio}
%\end{eqnarray*}
reported a $ 2.5~\sigma$ deviation from the standard model
(SM) prediction, $R_K^{SM}=1.0003\pm0.0001$ \cite{Bobeth:2007dw,Bordone:2016gaq}.
Similar trends were observed in the measurement of $R_{K^*}=\mathcal{B}(B^0\rightarrow K^{*0}\mu^+\mu^-)/\mathcal{B}(B^0\rightarrow K^{*0}e^+e^-)$  \cite{Aaij:2017vbb}. Both the measurements  indicate  LFUV in neutral current decays: a sign of new physics.

A simple way to quantify these observations is to consider model  independent fits to Wilson coefficients of the following effective operators:
\begin{equation}
	\mathcal{H}_{eff}=-\frac{G_f\alpha}{\sqrt{2}\pi}V_{tb}V^*_{ts}\sum_iC^i_{XY}\mathcal{O}^i_{XY}
	\label{effhamiltonian1}
\end{equation}
where the operator $\mathcal{O}^i_{XY}=(\bar s\gamma_\mu P_X b)(\bar l\gamma^\mu P_Y l)$ and $C_{XY}$ are the corresponding Wilson coefficients defined as $C_{XY}=C^{SM}_{XY}+C_{XY}^{NP}$.
There are several analyses involving different combinations of Wilson coefficients (WC) for the leptons. \cite{Hurth:2014vma,Hurth:2016fbr,Aebischer:2019mlg,Alok:2019ufo,Alguero:2019ptt,DAmico:2017mtc,Kumar:2019qbv,Ciuchini:2019usw}.

In this letter we make the first attempt to study the implications from low energy parity violation  experiments on these fits.  
Particularly, we are interested in the recent measurements of the weak charge of the proton $Q^p_W$ \cite{Androic:2018kni} and the  Caesium atom $(Q^{Cs}_W)$. 
These experiments  correspond to  a measurement of  axial-electron-vector-quark weak coupling constants $C_{1q}$ defined as: 
\begin{equation}
\mathcal{L}_{Q_W,Q_P}=\frac{\bar e\gamma_\mu\gamma_5 e}{2v^2}\sum_{q=u,d}C_{1q}\bar q\gamma^\mu q
\label{SM}
\end{equation}
The main goal of this paper would be to limit the extent of the electron contribution to the anomalies by means of these precise measurements. This conclusions drawn on the model parameters would serve as a hard upper-bound for a given class of NP models.

\begin{figure}[h!]
 	\begin{center}
 		\begin{tabular}{c}
 			\includegraphics[width=4.2cm,height=3.4cm]{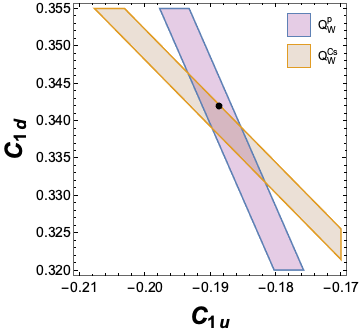}
 		\end{tabular}
 	\end{center}
 	\caption{ Allowed regions in the $C_{1u}$$C_{1d}$ plane due to measurements of weak charge of proton (gray) and Caesium (brown). The central value in the SM is represented by the black point.}
 	\protect\label{constraint}
 \end{figure}

The tree-level expressions for $C_{1q}$ in the SM are given as:
$C_{1u}=-\frac{1}{2}+\frac{4}{3}\sin^2_{\theta_W}, C_{1d}=\frac{1}{2}-\frac{2}{3}\sin^2_{\theta_W}$
In the SM, the values for $C_{1q}$ are: $C^{SM}_{1u}=-0.1887\pm 0.0022$ and $C^{SM}_{1d}=0.3419\pm0.0025$.
 The expressions for the weak charge of the proton and Caesium atom (in terms of $C_{1q}$) are given as:
\begin{eqnarray}
Q_W^p&=&-2\left[2C_{1u}+C_{1d}\right]\nonumber\\Q_W^{Cs}&=&- 2\left[55(2C_{1u}+C_{1d})+78(C_{1u}+2C_{1d})\right]
\label{weak}
\end{eqnarray}
Following the independent measurements for the proton \cite{Androic:2018kni} and the Caesium atom \cite{Dzuba:2012kx}, the allowed ranges at 1$\sigma$ are:
\begin{equation}
Q_W^p=0.0719\pm0.0045\;\;;\;\;Q_W^{Cs}=- 72.58(29)_{expt}(32)_{theory}
\label{weakcharge}
\end{equation}
 Fig.~\ref{constraint}  illustrates the simultaneous compatibility of both these measurements, showing the 2$\sigma$ ranges allowed by the measurement of weak charge of proton (gray) and Caesium (brown) in the $C_{1u}-C_{1d}$ plane. The black point represents the SM central value  and and lies in the region of overlap due to both  experiments. 

\subsection{New physics contributions}
Any NP contribution to either the $C_{1u}$ or $C_{1d}$ must satisfy the constraints from both the measurements simultaneously and will be the focus of the following discussion.
The coefficients $C_{1q}$ in Eq.~\ref{SM} can receive corrections due to different extensions of the SM. They can be induced either  at tree level due to the direct exchange of heavy vectors or at one-loop.
A generic NP extension to Eq.~\ref{SM} is given as:
\begin{eqnarray}
\mathcal{L}=\frac{\bar e\gamma_\mu\gamma_5 e}{2v^2}\sum_{q=u,d}C^{eff}_{1q}\bar q\gamma^\mu q\nonumber
\label{eff}
\end{eqnarray} 	
where $C^{eff}_{1q}=C^{SM}_{1q}+C^{NP}_{1q}$ and correspondingly lead to corrections to Eq.~\ref{weak}. Similar to the SM, the $C^{NP}_{1q}$  can  be factored into the  NP axial vector coupling  to electrons ($g_e^{AV}$) and the vector coupling to light quarks ($g_q^{V}$)
%\abovedisplayshortskip=0pt
%\belowdisplayshortskip=0pt
%\abovedisplayskip=10pt
%\belowdisplayskip=10pt
\noindent and can be expressed as:
%\begin{equation}
$C^{'NP}_{1q}=g_q^{V}g_e^{AV}$.
%\label{atphy}
%\end{equation}
%\noindent 
 Fig.~\ref{constraint} shows the simultaneous region of compatibility, in the plane of $C_{1u}-C_{1d}$, due to the two parity violation experiments. NP to contributions to $C_{1q}$ and the corresponding constraints on the model parameters were considered for instance in
  \cite{Erler:2003yk,Bouchiat:2004sp,Falkowski:2017pss,Schmaltz:2018nls,Benavides:2018rgh}.
  
Given our discussion on the $B$ anomalies and low energy parity violation experiments, it is not unusual to expect independent contributions to them in a generic NP model. In this paper we consider the possibility of an interplay between the two.

 \subsection*{Anomalies to parity violation}
 While the anomalies correspond to a flavour changing observable, $Q_W^{Cs,p}$ deal with a flavour diagonal transition. Thus, a correlation is possible only with the aid of an underlying model characterized by a flavour symmetry.
  To facilitate this correlation, we consider the SM to be augmented with an additional heavy neutral vector $Z^\prime$. The effective Lagrangian (after EWSB), parametrizing its  couplings to the fermions is given as \cite{Gauld:2013qja}
  \begin{eqnarray}
  \mathcal{L}&=&\frac{Z^{'\mu}}{2\cos\theta_w} \left[ g_e(g'_e)\bar e\gamma_\mu P_{L(R)}e +g_\mu(g'_\mu)\bar \mu\gamma_\mu P_{L(R)}\mu   \right. \nonumber\\&+&\left. \sum_{q}(g_q\bar q\gamma_\mu P_Lq+g'_q\bar q\gamma_\mu P_Rq)\right.\nonumber\\&+& (g_t-g_q)V^*_{ts}V_{tb}\bar s\gamma_\mu P_{L,R}b  +\ldots \left. \right ] 
  \label{zp}
  \end{eqnarray}
%\\[12pt]
The Lagrangian in Eq. \ref{zp} is characterized by the following features:
\begin{itemize}
	\item The up quarks are assumed to be in the mass basis. The rotation matrix ($D$) in the down sector is thus $D=V_{CKM}$.
	\item $U(2)$ flavour symmetry in the $Z'$ coupling to the quarks. This is essential in obtaining a $V_{CKM}$ like scaling in order to satisfy the constraints from $\Delta F=2$ processes \cite{Isidori:2010kg,Gori:2016lga,Bona:2007vi,Bona:2016bvr,Bona:2017gut}.
	
	\item \textit{$Z$-$Z^{'}$ mixing:} The mixing could be induced by vacuum expectation value (vev), kinetic mixing or loop induced. Since we attempt to represent a wide category of $Z^\prime$ scenarios we assume a mass mixing of the form $c\frac{m_Z^2}{m_{Z^\prime}^2}$ where $c\sim \mathcal{O}(1)$.  
	$c=1$ gives a contribution to the $Z\rightarrow f\bar f$ coupling of the form $g_f\frac{m_z^2}{m^2_{Z^\prime}}$, where $g_f$ is the $Z^\prime$-$f\bar f$ coupling. 
	The constraint on the size of $g_f$ for the leptons from $Z-Z'$ mixing is particularly strong    which translate into an upper bound on $g_f$ as $g_f\frac{m_Z^2}{m_{Z^\prime}^2}\lessapprox 0.001$.  These bounds can be relaxed with a custodial symmetry with custodial fermions  \cite{Blanke:2008yr,Blanke:2008zb,DAmbrosio:2017wis}. Alternately, this mixing could be also induced at loop level \cite{Gauld:2013qja} or a kinematic mixing with a small mixing parameter \cite{Davoudiasl:2012ag} enabling a relaxation of the constraints. 
	In order to represent a significant fraction of $Z^\prime$ scenarios, in this analysis we choose $g_f$ for the leptons such that  $g_f\frac{m_Z^2}{m_{Z^\prime}^2}$ is at-most $\sim 0.001$.
	
\end{itemize}
 For a study of the implications of these measurements on the  $B $ anomalies fits,  we will consider the  extreme possibility where the NP couples solely to the electron. While this assumption is extreme and specific,  it serves to address the following: At the moment, the verdict is yet to be out on the pattern of the $B$ anomalies in terms of the coupling of NP to electrons and muons. Instead of proceeding with an analysis involving both the leptons for the anomaly, we seek to answer to what extent can one go in terms of the coupling of the NP to the electrons. This would then serve as a hard upper-bound even in an analysis where both the leptons are involved. However, we also provide an insight on the impact of these measurement on the combined fits involving both the leptons.

 One dimensional fits involving electron. in the basis of Eq.~\ref{effhamiltonian1} were considered in \cite{DAmico:2017mtc} and will constitute the starting point for the analysis. The results of the fit for different combinations of quark and lepton (electron) chirality are given in Table. \ref{electrononly}. 
\begin{table}
	\begin{center}
		\begin{tabular}{| c | c | c | c |c|}
			\hline
			& operator & Best fit & 2 $\sigma$&$\sqrt{\chi^2-\chi^2_{SM}}$  \\ \hline
			Case A&($\bar s_L\gamma^\mu b_L$)($\bar e_L\gamma_\mu e_L$)&0.79&[0.29,1.29]&3.5\\\hline
			Case B	&($\bar s_L\gamma^\mu b_L$)($\bar e_R\gamma_\mu e_R$)&-3.31&[-4.41,-2.21]&3.8\\\hline
			Case C	&($\bar s_R\gamma^\mu b_R$)($\bar e_R\gamma_\mu e_R$)&-3.32&[-4.72,-1.92]&2.7\\\hline
		\end{tabular}
	\end{center} 
	\caption{2$\sigma$ ranges used for the fits to Wilson coefficients in the case where only electron couples to New Physics \cite{DAmico:2017mtc}.}
	\label{electrononly}
\end{table}

 We fit the Wilson coefficients for the anomalies at $B$ meson scale and determine the correlation between the different couplings. 
This correlation between the couplings $g_q, g_e$  is then used to compute its effects on the $C_{1q}$ which are determined at $q^2=0$.

In Table~\ref{electrononly}, we assume the WC due to the muon to be negligible. In addition we consider the dominance of one operator at a time.
 Corresponding to Table \ref{electrononly} we discuss each of the possibilities below:\\

\textbf{Case A} ($g_e'=0$): The lepton singlets are assumed to couple with a vanishing strength in Eq.~\ref{zp}. Assumption of a $U(3)$ symmetry in the coupling of the quark singlets to the $Z^\prime$ results in the absence of corresponding tree level FCNC.
Thus the most dominant NP operator contributing to $b \rightarrow sll$ is $\mathcal{O}_{LL}$ with the corresponding Wilson coefficient $C_{LL}$ given as 
\begin{equation}
C_{LL}=\frac{\sqrt{2}\pi g_e(g_t-g_q)}{4\cos^2\theta_Wm_{Z^\prime}^2 G_F\alpha}
\label{wc1}
\end{equation}
To extract the corrections to $C_{1q}$, just two quantities are required: vector coupling of light quarks ($g_V^q$) and axial vector coupling $g^{AV}_e$ of electron to $Z^\prime $. 
Assuming a $L\leftrightarrow R$ symmetry in the coupling of light quarks to $Z^\prime$, we get  $g_V^q=g_q$. On the other hand, the axial vector coupling of the electron is simply $g^{AV}_e=g_e/2$. Using this, the coefficients $C_{1q}$ get corrected as 
\begin{equation}
C^{eff}_{1q}=C^{SM}_{1q}+\frac{2v^2g_eg_q}{8\cos^2\theta_W m_{Z^\prime}^2}
\end{equation}

To represent a wide category of $Z^\prime$ models, we choose a broad range for the values of the fermion couplings:
\begin{equation}
g_e\in[0.02,2]\;\;\;g_t\in[0.02,2]\;\;\;g_q\in[0.02,2]
\end{equation}
While the ranges are general, care is taken to be consistent with different precision data. 
 The upper bound for the other coupling is chosen  so as to be consistent with an $\sim\mathcal{O}(1)$ parametrisation as well as being within the perturbativity bound of $g^2/4\pi \leq 1$. As we shall see below the results do not depend on the upper limit of the numerical scans.
 
\begin{figure}[htb!]
	\begin{center}
		\begin{tabular}{c}
			\includegraphics[width=4.cm,height=3.4cm]{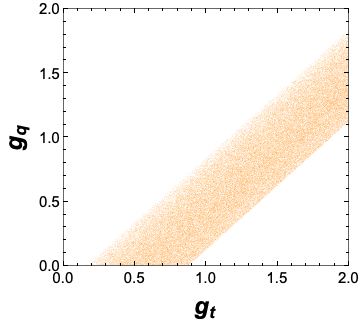}	\includegraphics[width=4.4cm,height=3.6cm]{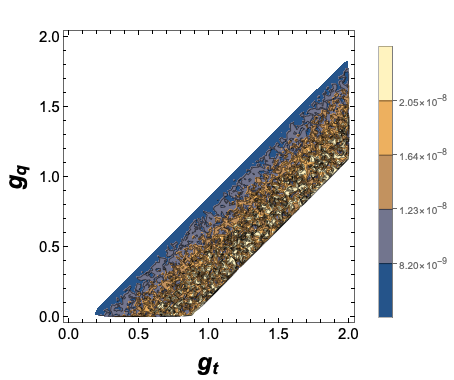}
		\end{tabular}
	\end{center}
	\caption{ Left plot gives the correlation in the $g_t-g_q$ plane for the anomalies. The right plot gives contours (in GeV$^{-2}$) extracted from the anomaly coefficient as a function of $g_q,g_t$.}
	\protect\label{p1}
\end{figure}

 Note that the coefficient of the effective operator in   Eq. \ref{eff} contributing to $C_{1q}$ is simply $ g_eg_q/M^2_{NP} $.
 The combined measurements in \cite{Androic:2018kni} lead to an upper bound on $M_{NP}/(\sqrt{g_eg_q}) > 8.4$ TeV. 
 Thus, the scale can be inverted to get the upper bound on this  coefficient of the four-fermi operator to be $1/8.4^2\sim 1.41\times 10^{-8}$ GeV$^{-2}$. 
 Now consider  the coefficient of the four fermi operators contributing to the anomalies. From the best fit for the  Wilson coefficients $C_{LL}\simeq 1$, we get $C_{LL}\frac{G_F\alpha}{\sqrt{2}\pi} \sim 10^{-8}$ GeV$^{-2}$.  Thus the four-fermi operators corresponding to both the parity violation experiments and the anomalies have a similar sensitivity to NP. This would ordinarily imply that parity violation experiments would not have a drastic effect on the solution to the anomalies.
 However, it is interesting to note the implications of this observation on the individual couplings of the $Z'$ to the fermions.
  %For a $Z'$ model, in addition to the mass of the heavy neutral vector, there are three additional parameters: $g_q,g_t,g_e$. The Wilson coefficient for the anomaly is proportional to $g_e(g_t-g_q)$.
 There are two things to be considered at this point:
\begin{itemize}
	\item The size of the difference $(g_t-g_q)$ in Eq. \ref{wc1} is independent of the individual sizes of $g_t$ and $g_q$. The only observable constraining the size of $(g_t-g_q)$ are the $\Delta F=2$ observable. Thus, a small $g_t$ and $g_q$ would give the same effect as a comparatively larger $g_t$ and $g_q$. This is clearly illustrated in the left plot of Fig. \ref{p1}. 
	\item In the same plane  of $g_t,g_q$, the right plot gives contours of $C_{LL}\frac{G_F\alpha}{\sqrt{2}\pi}$ (in GeV$^{-2}$) extracted  from the four-fermi operators for the anomaly . Note that the contours have the same order of magnitude which is expected for the explanation of the anomaly. The difference in the values of the contours corresponds to the $2\sigma$ range for $C_{LL}$ which are a function of $g_q,g_e,g_t$. As expected there is no explicit dependence on $g_t,g_q$. 
\end{itemize}

\begin{figure}[htb!]
	\begin{center}
		\begin{tabular}{c}
			\includegraphics[width=4.2cm]{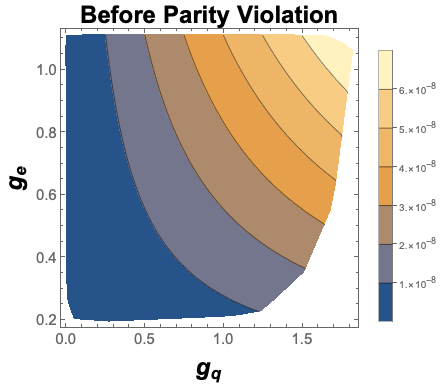}	\includegraphics[width=4.2cm]{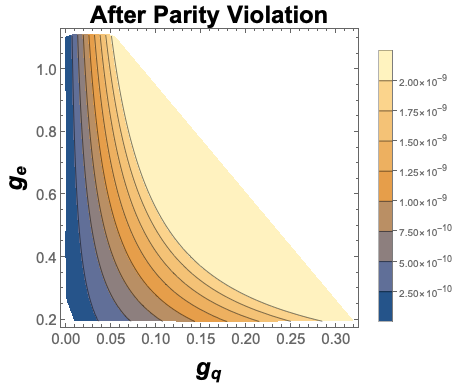}
		\end{tabular}
	\end{center}
	\caption{ Contours of $g_eg_q/M^2_{Z'}$ in the $g_e-g_q$ plane. Left plot corresponds to all values of $g_e,g_q$ which satisfy the $B$ anomaly while the right plot is limited to constraints from parity violation experiments..}
	\protect\label{p2}
\end{figure}

Now we move to the computation of $g^2/\Lambda^2$ from the parity violation experiments. Note here $g^2=g_eg_q$. Using the values which satisfy the anomaly, we plot contours of $g_eg_q/M^2_{Z'}$ in the $g_e-g_q $ plane in Fig. \ref{p2}. Note the difference between the left and the right plot. While the parameter space of $g_e$ is unaffected, the corresponding range of $g_q$ changes on account of the imposition of the parity violation constraints. This effect on the couplings due to the parity violation experiments is mainly due to the fact that the corresponding coefficient is bi-linear in $g_eg_q$. Thus a large $g_q$ (and correspondingly $g_t$) is prohibited.

\begin{figure}[htb!]
	\begin{center}
		\includegraphics[width=4.2cm]{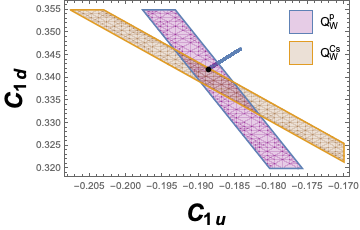}\includegraphics[width=4.1cm]{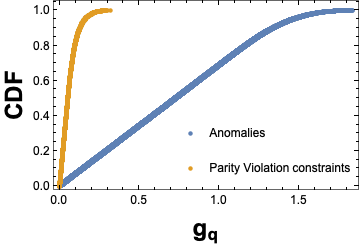}
	\end{center}
	\caption{Results with electron only fits for Case A. Left plot gives the projection on the $C_{1u}-C_{1d}$ plane, while the right plot represents the changes in the range for $g_q$. The blue (orange) curve represents the CDF before (after) the imposition of parity violation constraints.}
	\protect\label{constraint2}
\end{figure}

\begin{figure}[htb!]
	\begin{center}
	\includegraphics[width=4.2cm]{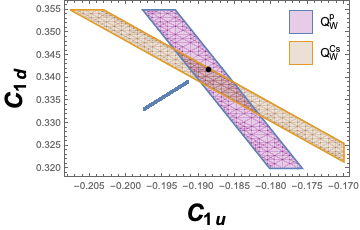}	\includegraphics[width=4.2cm]{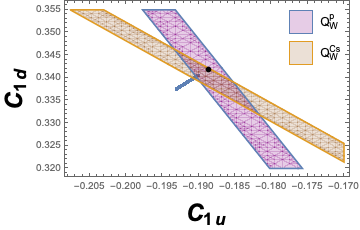}
%	\\\includegraphics[width=4.1cm]{range2.png}\includegraphics[width=4.1cm]{range2a.png}
	\end{center}
	\caption{Results with electron only fits for Case B.  Left plot corresponds to the case $g_q=g'_q$ while the  right corresponds to $g_q\ggg g'_q$. 
          }
	\protect\label{constraint3}
\end{figure}

A visual representation of the effect of the parity violation experiments on the solution to the anomalies is presented in the left plot of Fig.~\ref{constraint2}. 
Note that a only a small subset of the the solutions is admissible by the constraints from parity violation experiments. As shown in the right plot of Fig. \ref{p2}, the parameter space of the light quark coupling $g_q$ (and correspondingly $g_t$) is affected.  
This can be further quantified by the definition of the following variable:
  \begin{equation}
  \text{CDF}(x)=\int_{-\infty}^x\mathcal{P}(x)dx
  \end{equation}
This can be understood as follows: Corresponding to  a given set of solutions ${X}$,
 for any given point on the $x$ axis, the CDF expresses the percentage of solutions in ${X}$ such that $X\leq x$. $CDF=1$ for a given $x_a$ implies all solutions satisfy $X\leq x_a$.
  Right plot of Fig.~\ref{constraint2} gives the CDF for the light quark coupling $g_q$.  The uniform increase in the CDF for the blue curve is indicative of the fact that the   range $[0,1.8]$ is admissible. The red curve on the other hand corresponds to the case when the limits from parity violation  experiments are imposed. It rises rapidly and reaches $\sim 1$ at around $g_q\sim 0.23$ which corresponds to the maximum allowed value. Note that the case $g_q\rightarrow 0$ is admitted by both the anomaly solutions as well as the parity violation data. It represents the limiting case $C^{eff}_{1q}\simeq C^{SM}_{1q}$.
  This bound will have implications for the direct production cross sections and will be discussed later
 \\
\textbf{Case B} $g_e=0$:We now consider the other extreme possibility where only the electron singlets couple to $Z^\prime$.
As seen in Table \ref{electrononly}, this case gives the best fit of the three cases considered.
 The Wilson coefficient in this case is given as:
\begin{equation}
C_{LR}=\frac{\sqrt{2}\pi g'_e(g_t-g_q)}{4\cos^2\theta_Wm_{Z^\prime}^2 G_F\alpha}
\label{wc2}
\end{equation}
From the fits in Table \ref{electrononly}, it is important to note that the sign is reversed relative to Case A. Considering an implementation of the coupling ranges similar to Case A,  the negative value is only possible for $g_q>g_t$. For the estimation of corrections to $C_{1q}$, it is important to note that unlike the earlier case, only the right handed electron current couples to new physics. Thus,  corresponding axial vector electron current coupling is simply $g^{AV}_e=-g'_e/2$. For the light quark case we first begin with the assumption of  $L\leftrightarrow R$ symmetry: $g_q=g'_q$.
Thus, the coefficients $C_{1q}$ are given as as 
\begin{equation}
C_{1q}=C^{SM}_{1q}-\frac{2v^2g'_eg_q}{8\cos^2\theta_Wm_{Z^\prime}^2} 
\end{equation}
The results are illustrated in the top left plot of Fig.~\ref{constraint3}.
Unlike Case A, the limiting case does not reduce to the SM as seen in the top left plot of Fig \ref{constraint3}. This is a consequence of the fact that for the solutions to the anomalies, the Wilson coefficients are negative. They are proportional to $(g_t-g_q)$ where $g_q>g_t$. Thus $g_q\rightarrow 0$ is not permitted. However, these solutions are not compatible with the constraints from low-energy physics. Thus the best fit scenario as per Table \ref{electrononly} is not admissible in a simple realization of $Z^\prime$.
The large contributions to $C_{1q}$ are due to the relative strength of $g_q$ and the fact that $g_q^V=g_q$. If we assume $g_q\ggg g'_q$ \textit{i.e.} no $L\rightarrow R$ symmetry, then $g_q^V=g_q/2$
 and the coefficients $C_{1q}$  now get corrected as: 
\begin{equation}
C_{1q}=C^{SM}_{1q}-\frac{2v^2g'_eg_q}{16\cos^2\theta_Wm_{Z^\prime}^2} 
\end{equation}
The corresponding results are now shown in the top right plot of Fig.~\ref{constraint3}. The agreement with the constraints from $Q^p_W$ and $Q^{Cs}_W$ is due to the reduction of the numerical value of $g_q^V$ with respect to the case with $L\leftrightarrow R$ symmetry for the light quarks. A coupling structure of this form can arranged for instance in a warped framework where the doublets are more composite than the singlets.
 It has to be noted in this case that the minimum right handed electron coupling to $Z^\prime$ ($g'_e $) is at the edge of $g'_e\frac{m^2_Z}{m^2_{Z^\prime}}\simeq0.001$.  A minor departure from $\mathcal{O}(1)$ mixing would easily accommodate such couplings \cite{DAmbrosio:2019tph}.
 
 \textbf{Case C} : This case is characterized by the presence of tree level FCNC due to the non-universality in the coupling of the quark singlets.  The doublets are assumed to couple universally to $Z^\prime$. With the assumption that the right handed rotation matrix $D_R$ has a $V_{CKM}$ like structure, then this case is numerically similar to that of \textbf{B}. The only difference being that a requirement of consistency with parity violation data would necessitate $g'_q\ggg g_q$ in this case.

\section{Collider Implications}
\label{collider}
\begin{figure}[htb!]
	\begin{center}
		\begin{tabular}{cc}
			\includegraphics[width=4.2cm]{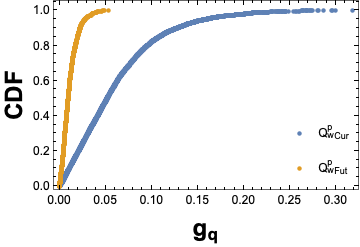}&\includegraphics[width=4.2cm]{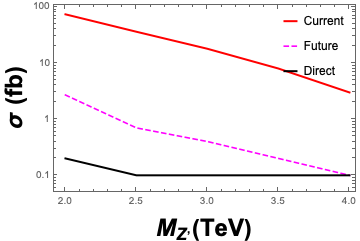}\\\includegraphics[width=4.2cm]{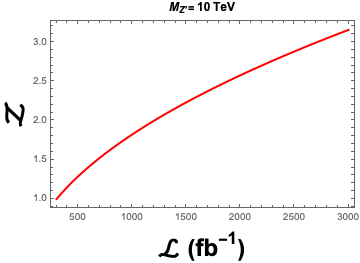}&\includegraphics[width=4.6cm,height=2.8cm]{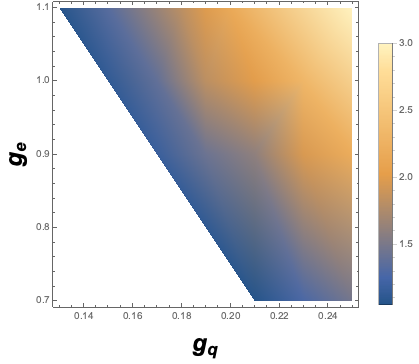}
		\end{tabular}
	\end{center}
	\caption{ Top: left plot gives the change in $g_q$ with $10\%$  improvement in  $Q_W^{Cs}$ for a 3 TeV mass. Right plot gives the computation of the maximum allowed cross section  with current (red) and after (pink dotted) improvement of $Q_W^{Cs}$. Bottom: The plots illustrate the reach for a 10 TeV resonance. The left gives the signal sensitivity as a function of the integrated luminosity using the current bounds for parity-violating experiments. The right  plot gives the corresponding regions in the $g_q-g_e$ space that can be probed at 3 $ab^{-1}$ with shading representing the corresponding sensitivity.}
	\protect\label{constraint4}
\end{figure}

The solutions consistent with the constraints from low-energy parity violation also have an interplay with direct searches. We discuss this correlation in the context of the cases discussed above:\\
\textbf{Case A:} In this scenario, the limiting case being the SM ($g_q\rightarrow 0$), strong upper bounds were obtained on the value of $g_q$ and hence the $Z^\prime$
production cross section. It will be interesting to compare it with the
bound from direct searches.
In a $Z^\prime$ model with coupling to electrons, there exists strong limits from direct searches for both ATLAS \cite{Aad:2014cka} and CMS \cite{CMS:2016abv}. For instance for a 3 TeV resonance decaying into $ee$, there is an upper bound on $\sigma\times B.R. <0.5$ fb. Since the solutions to the anomalies correlate the coupling of the light quarks to those of the third generation as well as leptons, upper bounds on $g_q$ and correspondingly $\sigma_{Z^\prime}$ can be obtained and compared with those obtained from direct searches. In the first instance, we assume $100\%$ branching fraction in the electrons. Consider the case where the coupling to the muons is zero. Left plot of Fig.~\ref{constraint4} gives the change in  the magnitude of light quark coupling for a 3 TeV resonance with the current (blue) and with $10\%$ improvement in the measurement of $Q_W^{Cs}$. The corresponding changes in the cross sections are given in the right plot of  Fig.~\ref{constraint4} for different masses which give the upper bound on $\sigma_{Z^\prime}$ for different benchmark masses.
The upper bound on $\sigma\times B (Z^\prime\rightarrow ee)$ for the corresponding masses is given by the solid black line corresponding to the values extracted from \cite{Aad:2014cka} \footnote{For a 4 TeV resonance we assume $\sigma B<1$ fb.}. 
The significance of this result lies in the fact that even with an unrealistic assumption of $100\%$ branching fraction into electrons, a mild improvement in the APV sensitivity could be comparable with the bounds from direct searches . If one assumes a SM like branching fraction of $3\%$ into electrons, the current sensitivity is roughly compatible with the direct searches for   masses 3 TeV and higher.
Improvements by $\sim 10\%$ is illustrated by the dotted pink line thereby resulting in even better sensitivities. The upper-bound on the computed cross-section ($\sigma_{Z'}$) for masses $\geq 3.9$ TeV is better than the those obtained from direct searches where the bounds are computed on the variable $\sigma \mathcal{B}(Z^\prime\rightarrow ee)$. Thus in a given model with a known $\mathcal{B}(Z\rightarrow ee)$,  the bound from $Q^{Cs,p}_W$ can accommodate only  smaller values of $\sigma \mathcal{B}$ than those allowed by direct searches. 
LHC is also sensitive to probing non resonant NP effects by exploring event multiplicity at the tail of the di-lepton $p_T$ spectrum \cite{Greljo:2017vvb}. As an illustration we consider a non-resonant 10 TeV $Z^\prime$ production and explore the event multiplicities in the regime $p_T>900$ GeV. 
\footnote{The model file for the signal is  generated using {\tt{FEYNRULES}} \cite{Alloul:2013bka} and matrix element for the process is produced using {\tt{MADGRAPH}} \cite{Alwall:2014hca}. We use {\tt{PYTHIA 8}} \cite{Sjostrand:2007gs} 
	for the showering and hadronization.}
Using the CMS card of {\tt{DELPHES 3}} \cite{deFavereau:2013fsa}, we extract events with two isolated electrons, with the leading electron satisfying $p_T>900$ GeV. The events are then distributed into bins of size 100 GeV each and we compute the following variable \cite{Cowan}:
\begin{equation}
\mathcal{Z}=\sqrt{\sum_i\left(2(s_i+b_i)\log\left[1+\frac{s_i}{b_i}\right]-2s_i \right)}
\label{sensi}
\end{equation}
where $s_i(b_i)$ are the signal(background) events in the $i^{th}$ bin. The variable $Z$ is a measure of the signal sensitivity over the background expectation and is sensitive to the differences in events in individual bins. 
Bottom left plot of Fig.~\ref{constraint4} gives the signal discovery significance over the background as a function of the integrated luminosity. It clearly illustrates the sensitivity of the LHC in probing the tail of the $p_T$ distribution. The right plot gives the parameter space of couplings that can be probed at 3 $ab^{-1}$. The shaded regions indicate the corresponding signal sensitivity. A correlation between parity-violation physics and such indirect signatures would be interesting as a future exercise. 

Constraints from low-energy physics also have implications for the partial decay width of the $Z^\prime$ into fermions. Note that in most models $Z^\prime\rightarrow t\bar t$ constitutes the most likely channel for discovery. As shown in the top bottom right plot of Fig.~\ref{constraint3}, the allowed top-quark coupling to $Z^\prime$ also reduces after the bounds from weak charge measurements. It is to be noted that the electron coupling does not change drastically before and after the imposition of the parity-violating  constraints. After the imposition of the latter, the branching fraction into $t\bar t$ becomes comparable to that of $ee$.

\textbf{Case B(and C:)} This scenario is distinctly different from Case A owing to the opposite sign of the Wilson coefficients as required by the B anomalies. As a result the coupling to the light quarks $g_q$ is always greater that $g_t$. 
In this case the features of the branching fraction into a top-quark
pair can be classified into the following two categories:\\
1)$L\leftrightarrow R$ symmetry for the top couplings: With  the assumption of $g^t_L=g^t_R$ and with $g_e>g^t_{L,R}$, will result in a comparatively lower branching fraction into a top-quark pair. Thus leptons are likely to constitute the most likely discovery mode.\\
2)$g^t_L<g^t_R$ for Case B: Since the flavour diagonal coupling  of the top singlets is a free parameter, one can also accommodate a larger value of its coupling. This results in the possibility of a larger branching fraction compared to scenario 1.  A similar argument also applies to Case C with the difference that the coupling of the top doublets is a free parameter and one can accommodate $g^t_L>g^t_R$. A large deviation between the coupling of the two  chiralities will also result in a forward backward asymmetry. 
However, updated analysis from  TEVATRON \cite{Aaltonen:2017efp} would strongly disfavour this scenario.

\vspace{-3mm}
\section{Impact of the muon}
\begin{figure}[htb!]
	\begin{center}
		\begin{tabular}{c}
			\includegraphics[width=4.2cm]{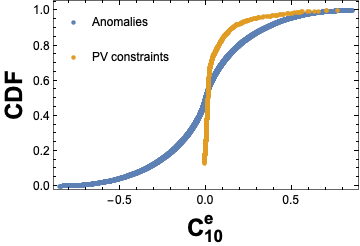}	\includegraphics[width=4.2cm]{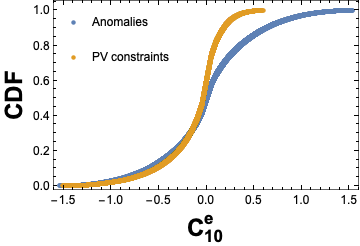}
		\end{tabular}
	\end{center}
	\caption{ The blue (orange) curve represents the CDF for $C^e_{10}$ before (after) the imposition of parity violation constraints for the case with left handed electron current (left) and right handed electron current (right)}
	\protect\label{Ce10}
\end{figure}

Thus far we have discussed the implications of the parity violation experiments on the fits involving only the electrons. This facilitated a direct interplay between the two sectors enabling us to draw significant conclusions on concluding about the validity of electron only solutions to the anomalies.
However, the strong constraints on the model with the electron only solutions strongly suggest the addition of the muon contribution to the anomalies. 
These measurements  could also impact the muon sectors in scenarios where the fits to the anomalies involve both the electron and the muon. In order for the parity violation experiments to have implications on the combined fits, they must involve the operator $C^e_{10}$. The simplest possibility is the four dimensional fit considered in \cite{Hurth:2016fbr} which includes  $C^{e,\mu}_{9},C^{e,\mu}_{10}$. Limits on the range of $C^e_{10}$ which has direct implications on the anomaly will also affect the corresponding ranges for the other operators. The ranges at $2\sigma$ for the combined fits is given below \cite{Hurth:2016fbr}:
\begin{eqnarray}
C_9^{\mu}/C^{SM}_9\in[-0.33,0.06]\;\;\;C_9^{e}/C^{SM}_9\in[-2.23,0.74]\nonumber\\C_{10}^{\mu}/C^{SM}_{10}\in[-0.29,0.14]\;\;\;C_{10}^{e}/C^{SM}_{10}\in[-2.60,0.60]\nonumber\\
\label{ranges}
\end{eqnarray}
where $C^{10}_{SM}=-4.103~ C^{9}_{SM}=4.211$.
We begin with the case where only left handed lepton currents are involved.
For the model under consideration,  $C^e_{10}$ is simply related to $C_{LL}$ as
$C^e_{10}=-C_{LL}/2$. From Eq. \ref{ranges}, the range of  $C^e_{10}$ consistent with the explanation for the anomaly in Eq. \ref{ranges}, is $-2.46<C^e_{10}<10.66$. Left plot of Fig.\ref{Ce10} gives the changes in $C^e_{10}$  for the model under consideration before and after the imposition of parity violation constraints. The blue curve is within the acceptable range in Eq. \ref{ranges}, while the range after imposition of parity violation (PV) constraints is reduced further. The lower negative bound on $C^e_{10}$ corresponds to the case where $g_q>g_t$ and correspondingly larger values of the difference $(g_q-g_t)$ are forbidden. 

This directly impacts patterns in the correlations between $C^e_{10}-C^\mu_{10}$, $C^e_{10}-C^e_{9}$   ,$C^e_{10}-C^\mu_{9}$. While it may not change the ranges of the muon operators it changes the pattern of the fits.

The bounds corresponding to cases with only right handed currents are different. This case is characterized by negative value of $C^e_{10}$ relative to the first case.
The right plot of Fig. \ref{Ce10} gives the corresponding changes in $C^e_{10}$ for this case.
In this case as well the  upper bound on the value of $C^e_{10}$ corresponds to the scenario where $(g_q-g_t)$. An important advantage of including muons in the fits to the anomalies is the  admissibility of $L\leftrightarrow R$ symmetry in the light quark couplings. As seen in Fig. \ref{constraint3} this symmetry needed to be broken to reduce the contribution to $C_{1u,1d}$. Similar to the earlier case, the constraint on $C^e_{10}$ will affect patterns in the  $C^e_{10}-C^\mu_{10}$, $C^e_{10}-C^e_{9}$   ,$C^e_{10}-C^\mu_{9}$ space. However, scenarios with right handed electron currents can be more viable with the assumption of contribution of both leptons to the anomalies.

\section*{Conclusions}
Anomalies in the semi-leptonic decays of the $B$ constitutes one of the strongest hints for non standard physics. It can be reconciled with fits to the effective theory by considering different patters of coupling of the leptons to the NP. Focusing on the extreme possibility involving only the electrons, we attempt the  first study of its correlation with low energy parity violation data. Working in a minimal model of $Z^\prime$, we determine the model parameters to fits involving different chiralities of quark and lepton current. Note that irrespective of the chirality of the electron current used to explain the anomalies,  the solutions are only marginally consistent with the contraints from parity violation data.
The best fits scenario with $C_{LR}$ Wilson coefficients is not admissible in a simple realization of $Z'$ where the light quarks coupling respect $L\leftrightarrow R$ symmetry.
We demonstrate the improvement in the constraints due to future measurements in $Q_W^{Cs}$ and $Q_W^p$, thereby strongly motivating the future directions in this regard  \cite{Becker:2018ggl}. This  will not only serve as complementary bound to those from direct searches but also serve to constrain the parameter space corresponding to states beyond the realm of resonant production of the LHC. This study hence points towards the inference that an additional muon contribution is necessary  which ameliorates the constraints from the parity violating experiments. We observe that while the inclusion of the muon ensures a greater degree of consistency with these experiments the pattern of the correlation changes. A detailed four dimensional fit in this context will be reserved for a future study.

\section*{Acknowledgments}
We are thankful to David Armstrong for highlighting the difference between the different parity violation experiments.
 We are grateful to G. Isidori, F. Mahmoudi,  M. Nardecchia and L. Silvestrini for several useful comments and a careful reading of the manuscript. We are also grateful to F. Feruglio for useful discussions at the beginning of the project. 
 G.D. and  A.I. are supported in part by MIUR under Project No. 2015P5SBHT and by the INFN research initiative ENP.
\bibliography{biblio1}

\end{document}